\PassOptionsToPackage{table}{xcolor}
\documentclass[%
 reprint,
 amsmath,amssymb,
 aps,
]{revtex4-2}

\usepackage{graphicx}
\usepackage{dcolumn}
\usepackage{bm}
\usepackage{siunitx}
\usepackage{amsmath}
\usepackage[english]{babel}

\usepackage{svg}

\usepackage{xcolor}
\begin{document}
\preprint{APS/123-QED}
\twocolumngrid
\title{Towards Navigation-Grade and Deployable Optomechanical Accelerometry}

 \author{Chang Ge$^1$}
 \author{Daniel Dominguez$^2$}
 \author{Allison Rubenok$^1$} 
 \author{Michael Miller$^2$}
 \author{Matt Eichenfield$^{1,2}$}\email{eichenfield@arizona.edu}

\affiliation{\phantom{$^1$ James C. Wyant College of Optical Sciences, University of Arizona, Tucson, Arizona, USA} \\ $^1$ Wyant College of Optical Sciences, University of Arizona, Tucson, Arizona, USA \\ $^2$ Microsystems Engineering, Science, and Applications, Sandia National Laboratories, Albuquerque, New Mexico, USA
}


\begin{abstract}
We design and experimentally demonstrate an architecture for achieving navigation-grade, fiber-packaged optomechanical accelerometers that can operate with a large dynamic range, over a wide temperature range, and without sophisticated laser sources. Our accelerometer architecture is based on a novel set of design principles that take advantage of the strengths of optomechanical accelerometry while eliminating many of its historical weaknesses. Displacement readout is provided by an integrated, differential strain-sensing Mach-Zehnder interferometer (DSMZI) attached to an ultra-rigid, bulk-micromachined proof mass having a 93.4 kHz fundamental resonance frequency (22.5 pm/g displacement). Despite the extreme rigidity, the high displacement sensitivity provides an insertion loss limited 4.2 $\mu g/\sqrt{\mathrm{Hz}}$ acceleration resolution, with a straight-forward path to achieving 330 $n g/\sqrt{\mathrm{Hz}}$ by improving the fiber-to-chip coupling. Further, we show that the combination of high rigidity and intrinsic differential optical readout makes the device insensitive to the common causes of bias instability, and we measure a bias instability of 6.3 $\mu g$ at 243 seconds. The DSMZI provides a 17 nm optical bandwidth and a temperature operating range of greater than 20 $^\circ\mathrm{C}$, both orders of magnitude larger than previous demonstrations of optomechanical accelerometers. The high rigidity and large optical bandwidth yield an expected dynamic range of 165.4 dB. The combination of high acceleration resolution, high dynamic range, low bias instability, and intrinsic insensitivity to wavelength, temperature, and package stresses makes our device well suited for deployment in realistic environments demanded by real-world applications and demonstrates a path for optomechanical accelerometers to ultimately exceed the performance of all other chip-based accelerometers.
\end{abstract}


\maketitle
\twocolumngrid
The state of the art in optomechanical inertial sensing has rapidly progressed since its first demonstration more than 10 years ago~\cite{krause2012high}. However, while benefiting from high sensitivity and small size, the sensors demonstrated thus far lack the ability to maintain precise operation for long periods of time in the environments required by many real-world applications. Navigation-grade accelerometers require performance in the regime of approximately 1 $\mu g/\sqrt{\mathrm{Hz}}$ of noise-equivalent accelerations and bias instabilities below 10 $\mu g$~\cite{vectornav_ins}. 
Prior demonstrations of optomechanical accelerometers have been primarily based on high-Q nano-optomechanical cavities, making them very sensitive to acceleration while simultaneously being intrinsically sensitive to fluctuations in operating wavelength, environmental temperature, and package stresses/body forces, for example, in environments of high vibration. These resonance devices require a laser that can be tuned exactly into resonance with the cavity and complex laser locking schemes to extend the dynamic range, as well as precise temperature control. These control systems increase the size, complexity, power consumption, and cost of the sensor. All of these factors are significant barriers to deployment in real-world applications, where environmental temperature changes and large vibrations that exert substantial force are common~\cite{darpa2024horcrex}. Additionally, these high sensitivities to temperature, wavelength, and body forces also give rise to a bias instability on the order of tens to hundreds of $\mu g$ over short integration
times~\cite{li2018characterization,jin2022micro,huang2020chip,gerberding2015optomechanical}, thereby resulting in the need for frequent external updates, for example by GPS, to be used for inertial navigation systems. Previous optomechanical accelerometers have demonstrated a range of impressive resolutions and bandwidths by utilizing a variety of different designs, as outlined in Table \ref{Table1}, however, none of these systems have been able to reach the navigation grade threshold.

\begin{figure*}[htbp]
\centering
\includegraphics[width=5 in]{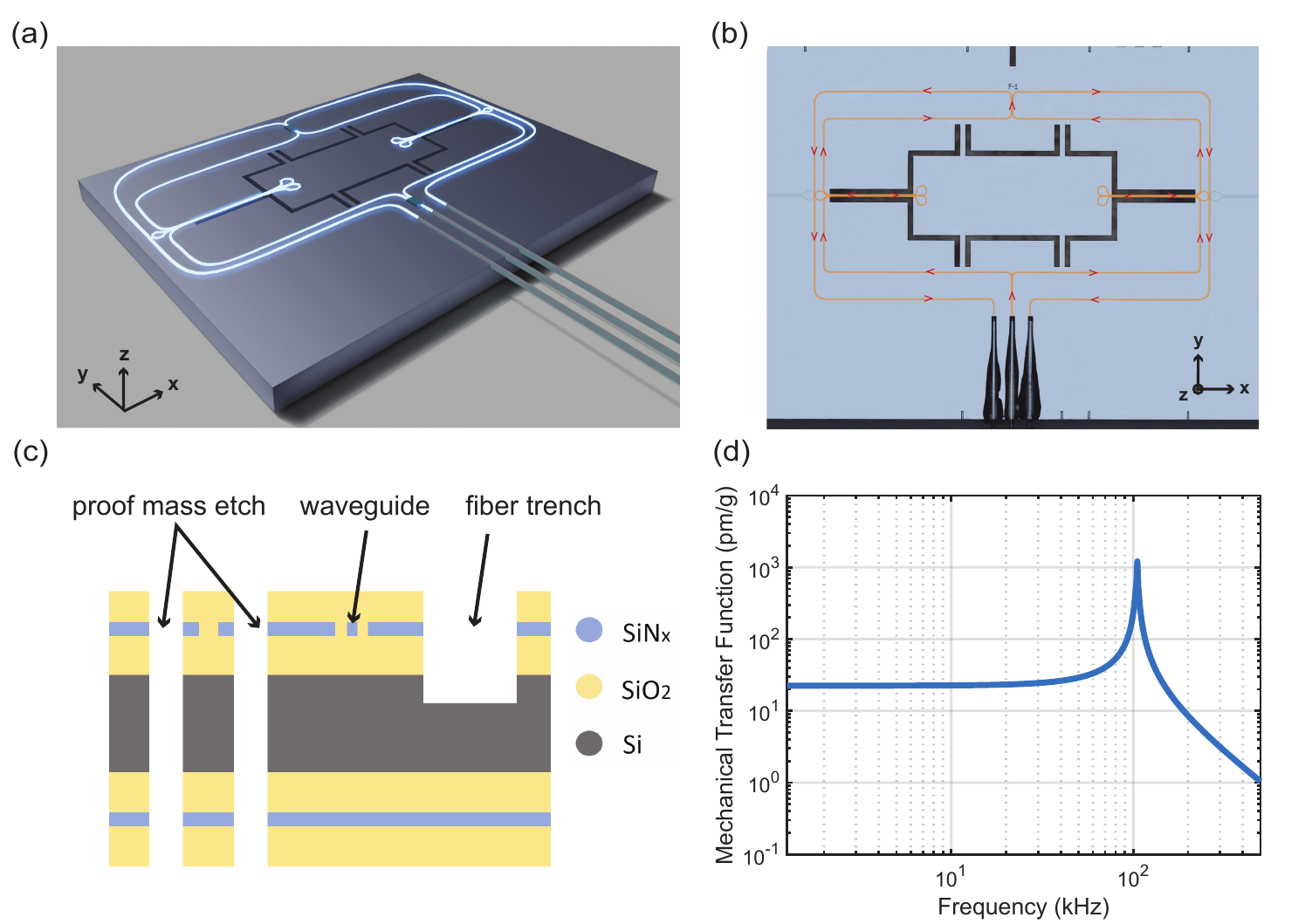}
\caption{\label{fig:fig1} (a) Illustrated schematic of the optomechanical accelerometer. (b) Microscope image of the fabricated 
 fiber-packaged accelerometer with the waveguides in false-color.  (c) CMOS compatible fabrication cross-section. (d) The mechanical transfer function derived from the equation with a mechanical Q of 54.}
\end{figure*}

Accelerometers operate by detecting the displacement of a proof mass relative to a reference frame, where that frame is attached to an accelerating body. The displacement of the proof mass, $x$($\Omega$), is proportional to the mechanical susceptibility:
\begin{equation} \label{eq:1}
\chi(\Omega) = \frac{x(\Omega)}{a(\Omega)} = \frac{1}{\Omega_s^2 - \Omega^2 + \frac{i \Omega_s \Omega}{Q}},
\end{equation} where $\Omega_s$ is the mechanical resonance frequency and $Q$ is its mechanical quality-factor. Below resonance, the displacement-to-acceleration sensitivity can be simplified to $dx/da$ $\approx$ 1/$\Omega_s^2$ = $1/(2\pi  f_s)^2$, where $f_s$ is the resonant frequency of the sense mode. This expression holds under the assumption that the frequency content of the applied acceleration is lower than the sense mode frequency and that the sense mode is the lowest frequency of the system that can be directly driven by acceleration or through nonlinear coupling from other directions. The displacement sensitivity, $S_{xx}$, determined by the particular readout method, represents the minimum detectable displacement per square root bandwidth. The sub-resonant acceleration sensitivity, where the response remains undistorted by resonant effects, is $S_{aa}$ = $S_{xx}$ ×$(\frac{dx}{da})^{-1}$ = $S_{xx}$(2$\pi$$f_s)^2$. Thus, for a given displacement sensitivity, $S_{xx}$, the only way to achieve higher acceleration sensitivity is to decrease the resonant frequency. However, decreasing the resonance frequency increases the displacement per unit acceleration below resonance as 1/$\Omega_s^2$. This reduces the sensor's linear range of operation and its ability to withstand large accelerations. Designing an inertial sensor that is robust to package stress caused by vibrations while maintaining high sensitivity requires that the accelerometer is substantially more sensitive to acceleration rather than body forces, $F_B$. This requires maximizing $\frac{dx}{da}$, while minimizing $\frac{dx}{dF_B}$. This optimization leads to optimizng the ratio:  
\begin{equation}  \label{eq:2}
\frac{dx}{da} \div \frac{dx}{dF_B} = \frac{1}{\Omega_s^2} \times k = \frac{1}{\Omega_s^2} \times m\Omega_s^2 = m.  
\end{equation}  
Thus, an accelerometer with a larger proof mass is inherently less susceptible to body forces and external vibrations, therefore reducing bias instability caused by those mechanisms. Bulk micromachining (BM) provides a robust approach for fabricating high-performance optomechanical accelerometers with monolithically integrated large proof masses. The BM technique utilizes deep etching through the full thickness of silicon substrates (500-700 $\mu m$), allowing the realization of a large-mass sensor platform (milligram scale).

Here we completely rethink how optomechanical accelerometers are designed, moving away from both small proof masses and high-Q cavities. Instead, we utilize a differential strain-sensing Mach-Zehnder interferometer (DSMZI) on the surface of a bulk-micromachined proof mass (BMPM). This architecture is intrinsically less susceptible to environmental temperature fluctuations, laser frequency noise, laser frequency drift, and package stresses/body forces \cite{dominguez2021megahertz,ge2024high}. We demonstrate a resolution as high as 4.2 $\mu g/\sqrt{\mathrm{Hz}}$ over a 3 dB operational bandwidth greater than 66 kHz, utilizing a 93.4 kHz fundamental resonance frequency (22.5 pm/g displacement). This is the largest operational bandwidth range reported in an optomechanical accelerometer to date, while still maintaining $\mu g/\sqrt{\mathrm{Hz}}$ sensitivity. The expected dynamic range of our device is 165.4 dB. We experimentally demonstrate that our accelerometer operates over an optical bandwidth of at least 17~nm and has a temperature operating range of greater than 20 $^\circ\mathrm{C}$. Furthermore, it allows a bias instability as low as 6.3~$\mu g$ at an integration time of 243 seconds, measured in air at room temperature, showing superior long-term stability compared to previously reported optomechanical MEMS devices~\cite{huang2020chip,li2018characterization,jin2022micro,gerberding2015optomechanical}. Given its resolution and frequency range, in addition to navigation, our accelerometer is well suited for applications including vibrometry, condition-based health monitoring, sports science, and impact or shock testing~\cite{murphy2020choosing,tandon2003detection,espinosa2015inertial}. 
\section{Theory} \label{sec:Design&Fab}

\begin{figure*}[htbp]
\centering
\includegraphics[width=5 in]{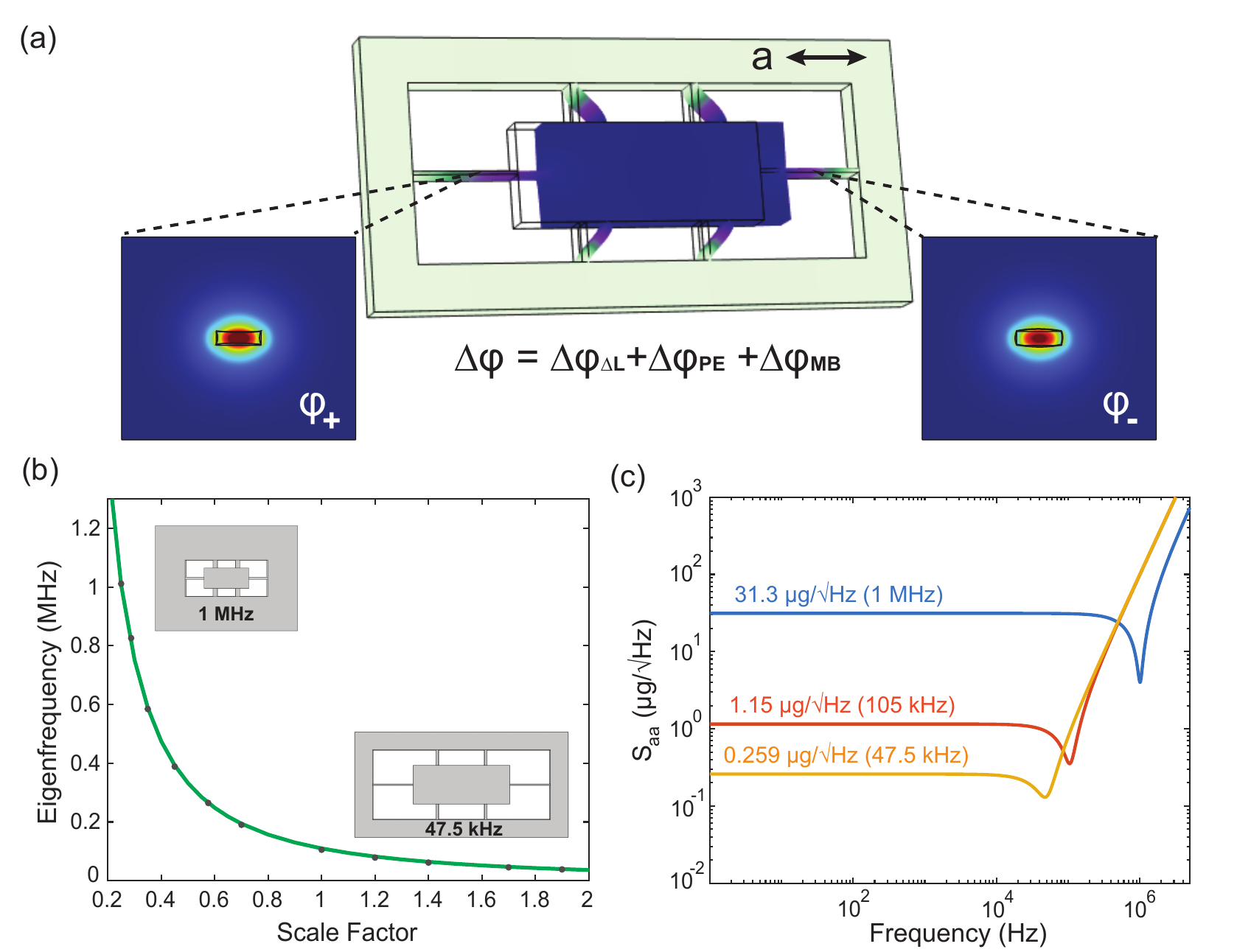}
\caption{\label{fig:fig2} (a) FEM model of the displacement field of the device at $f_m$, incorporating two waveguides with equal path lengths to form a balanced MZSI detection scheme. Differential deformation is optomechanically coupled to the waveguides via photoelastic and moving boundary effects. (b) A widely tunable bandwidth ($<$50 kHz to 1 MHz) achieved through geometric scaling of in-plane parameters. (c) Theoretical acceleration sensitivity for 47.5 kHz, 105 kHz and 1 MHz devices in air at room temperature.}
\end{figure*}

Our optomechanical accelerometer is depicted in Fig.~\ref{fig:fig1}a. A BMPM is suspended in air and attached to the accelerometer chip via six tethers. The device is designed to be sensitive along a single axis (x-axis). When subjected to acceleration along the x-axis, the two long, compliant tethers along the x-axis experience differential strain, while the four stabilizing tethers add stiffness to suppress undesired motion along the y-axis and torsional modes about all other axes. Our device is designed with a fundamental resonance frequency of $\Omega_m $ = 2$\pi$ × 105 kHz. As shown in Fig.~\ref{fig:fig1}d, the device is designed to operate in the flat frequency response region where $\Omega << \Omega_s$. The displacement of the proof mass per unit acceleration is given by $dx/da$ $\approx$ 1/$\Omega_s^2$ , which is equal to 22.5 pm/g.

The design of our optomechanical accelerometer aims to achieve a sensitivity comparable to a cavity-based optomechanical accelerometer, but with a much larger optical bandwidth. To this end, we employ a surface-integrated DSMZI. As shown in Fig.~\ref{fig:fig1}b, an integrated silicon nitride ($SiN_x$) waveguide on the surface of the silicon substrate is used to create an interferometer. The waveguide splits into two paths via a 1x2 multimode interference (MMI) coupler \cite{soldano1995optical,sheng2012compact}. After the MMI coupler, the light from each output is routed onto the x sensing tethers. After traversing the tethers, the waveguides loop back onto the sensing tethers towards the frame, where they loop back around returning to the proof mass one final time before looping back to the frame. This allows the optical path of the tether to be sampled a total of four times. After moving back onto the frame on the final pass, the light is then routed to a 2x2 MMI where it is recombined with the light that traversed the other tether.   


The DSMZI measures acceleration by detecting the displacement of the proof mass, which induces strain and deformation in the waveguides embedded in the tethers. To sense these changes, we analyze the transfer function of the optical power at the two output ports of the interferometer. One output port of the MZI produces an output optical power given by $P = P_{0}~l~\sin^{2}(\frac{\Delta\phi+\pi/2}{2})$, where $P_{0}$ is the input power, $l$ is the loss, and $\Delta\phi$ is the differential phase shift between the two arms. The change in phase induced by the strain and deformation under acceleration can be expressed as: 
\begin{equation} \label{eq:3}
\phi_{\pm} = \Delta(kL)_{\pm}(a) = k\Delta L(a)\pm \Delta (kL)_{PE}\pm \Delta (kL)_{MB}.
\end{equation}
The first term $k\Delta L(a)$ accounts for the change in the total optical path length within the strained tether. It is expressed as: 
\begin{equation} \label{eq:4}
k\Delta L(a) = k(L \pm \chi a),
\end{equation} 
where $k$ is the propagation constant, $L$ is the physical length of the waveguides, $\chi$ is the mechanical susceptibility ($dx/da$), and $\chi a$ is the displacement of the proof mass. Under zero acceleration ($a=0$), the optical phase shift $\phi$ simplifies to $kL$, resulting in $\Delta\phi = 0$. The second term, $\Delta (kL)_{PE}$, describes the change in the accumulated phase due to the photoelastic change in propagation constant, which arises from strain-induced modifications in the refractive index. Lastly, $\Delta (kL)_{MB}$ is the phase shift due to the opto-mechanical moving boundary effect, which induces a change in propagation constant due to waveguide dispersion with an altered cross-section by mechanical displacement \cite{johnson2002perturbation}.

The final differential phase, $\Delta\phi$, obtained by subtracting one arm’s accumulated phase by the other’s can be expressed as:
\begin{equation} \label{eq:7}
\Delta\phi = \phi_{+}-\phi_{-}=2[k\cdot (\chi a)+\Delta (kL)_{PE}+ \Delta (kL)_{MB}].
\end{equation}
A 3D finite element method (FEM) model is used to evaluate the integrals in Eq.(\ref{eq:SI kL_PE}, \ref{eq:SI kL_MB}) numerically using the exact strain profiles induced by acceleration of the proof mass. Fig.~\ref{fig:fig2}a shows the displacement field of the accelerometer when driven with a 1 g acceleration at a frequency substantially less than the resonance frequency. This causes an equal and opposite strain and deformation in the sensing tethers. The two inset figures in Fig.~\ref{fig:fig2}a depict the mode profiles in the oppositely deformed and strained waveguide cross sections, causing equal and opposite phase accumulation for the TE waveguide modes in each arm. To account for the photoelastically induced phase shift, $\Delta \phi$, described by Eq.(\ref{eq:SI kL_PE}), the refractive indices of the waveguide core (\(\text{SiN}_x\)) and the cladding (\(\text{SiO}_2\)) are modified by the photoelastic effect, given by 
{$\Delta\frac{1}{ n_{i}^2} = \sum_{j=1}^{6} p_{ij}S_{j}$\cite{chen2006appendix,tian2024piezoelectric}.} Here, \(p_{ij}\) is the photoelastic tensor and \(S_{j}\) is the strain tensor. 

This analytical model enables us to predict the sensor’s acceleration-to-power responsivity and noise-equivalent acceleration (NEA), while accounting for thermal, shot, and detector noise (see Appendices C \& D). Assuming 10 mW of optical power reaches each detector in a balanced detection scheme, the sensor is expected to achieve a displacement resolution of up to 0.03 fm$/\sqrt{\mathrm{Hz}}$ without the use of an optical resonator. This corresponds to an off-resonance, shot-noise-limited NEA of 1.15 $\mu g/\sqrt{\mathrm{Hz}}$, which is remarkable given the high stiffness of the mechanical structure. In addition, this model allows us to calculate the expected dynamic range, which we find to be 165.4 dB (see Appendix G). 

By modifying the proof mass size and tether dimensions, the sensor's resonance frequency can be tuned over a wide range, from $<$ 50 kHz to 1 MHz, as shown in Fig.~\ref{fig:fig2}b. This enables a wide range of applications with diverse performance requirements, including varying demands for resolution and bandwidth. We estimate with a 47.5 kHz design, the acceleration resolution would be as high as 0.259 $\mu g/\sqrt{\mathrm{Hz}}$. This would make these devices suitable for precise vibration monitoring \cite{murphy2017choosing,murphy2020choosing}, while a MHz-bandwidth design would provide a resolution of 31.3 $\mu g/\sqrt{\mathrm{Hz}}$, which would make it useful for shock testing \cite{tandon2003detection} (Fig.~\ref{fig:fig2}c).

\section{Results} \label{sec:ExperimentalResults}

\begin{figure*}[htbp]
\centering
\includegraphics[width=6.5 in]{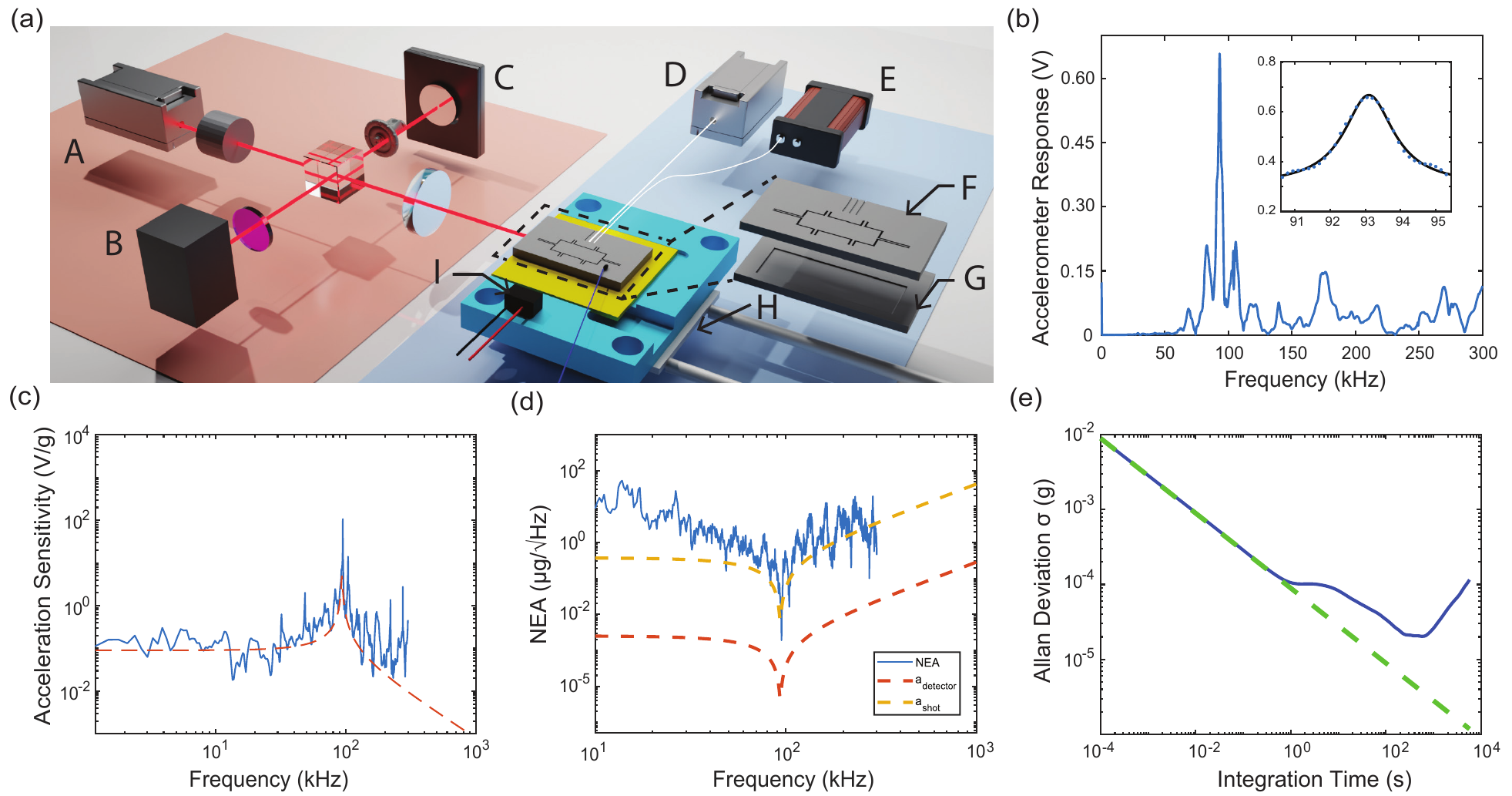}
\caption{\label{fig:fig3-Setup-Sensitivity-AllanDev} (a) Experimental Setup: A. 633 nm laser for free space reference interferometer. B. Photodiode for free space reference interferometer. C. Piezoelectically driven mirror. D. Tunable laser source 1490 nm - 1650 nm. E. Photodiode to detect optical output of chip accelerometer. F. Accelerometer Chip. G Mounting chip. H. Peltier Cooler. I. Temperature sensor. (b) Accelerometer response plot of the photodiode signal illustrating the mechanical resonance at 93.4 kHz. Inset: The response near resonance is fitted using the mechanical susceptibility, yielding a mechanical Q factor of 54.3. (c) Acceleration sensitivity as a function of frequency, with the dashed red line indicating the theoretical sensitivity derived from the analytical model. (d) Noise Equivalent Acceleration (NEA) versus frequency, showing the measured baseline noise floor. (e) Allan deviation of the device on a log-log scale, showing a minimum at 579.3 seconds, which corresponds to a bias instability of 20.3 $\mu g$. The green dotted line indicates the velocity random walk that corresponds to the white noise region, which appears as a slope of -0.5.}
\end{figure*}

Figure~\ref{fig:fig3-Setup-Sensitivity-AllanDev}a shows a schematic of the experimental setup. Our accelerometer chip (F) is mounted on a separate silicon chip (G) with a rectangular region etched out, allowing for free movement of the proof mass but rigid mounting of the frame. This assembly of accelerometer and mounting chip is then epoxied onto a shear piezo actuator and finally mounted atop a small metal holder. To enable temperature control, a temperature sensor (I) is affixed to the metal plate along with a Peltier cooler (H), placed underneath of it. A smaller temperature sensor is attached directly to the accelerometer chip to provide a more accurate temperature reading of the device itself. Light from a tunable laser (D) (1490 nm – 1650 nm), is coupled onto the chip via a fiber optic input aligned to the waveguide and epoxied into a deep-etched (60.5 $\mu$m) rectangular groove. 

The accelerometer is driven by a piezoelectric actuator using a sinusoidal wave generated by the internal source of a lock-in amplifier with a sinusoidal frequency sweep up to 300 kHz. This sinusoidal wave induces a displacement $x = b V_0 \sin(2 \pi ft)$, where $b$ is the piezo voltage-displacement constant, $V_0$ is the amplitude of the applied voltage and $f$ is the drive frequency. The resulting acceleration is given by $a = -x \omega_d^2$, where $\omega_d = 2\pi f$ is the angular frequency. The displacement of the accelerometer is calibrated using a free-space Michelson interferometer as a reference, as shown in Figure~\ref{fig:fig3-Setup-Sensitivity-AllanDev}a. Figure~\ref{fig:fig3-Setup-Sensitivity-AllanDev}b shows the photodiode output on the lock-in amplifier as a function of frequency, revealing a dominant peak at 93.4 kHz, which corresponds to the fundamental mechanical resonance. This closely aligns with the designed fundamental eigenfrequency of 105 kHz, with the difference likely being due to the angled sidewalls of the deep reactive ion etch (DRIE). The measured mechanical Q-factor in air is 54.3, thus with a mass of 16.7 mg, the thermal noise is given by $a_{th} = \sqrt{\frac{4 k_B T \omega_m}{m Q_m}}= 0.33$ $\mu g/\sqrt{\mathrm{Hz}}$. Figure~\ref{fig:fig3-Setup-Sensitivity-AllanDev}c shows the frequency-dependent acceleration sensitivity that is obtained by normalizing the photodiode signal with the applied acceleration in the drive frequency range. We show the measured acceleration sensitivity and plot the best fit (dotted red line) to this curve from Eq.~\ref{eq:1}. This allows us to extract dP/dA directly and compare it to the model from the previous section. We find a responsivity that is within a factor of 5.6 of the predicted model. The sensor's operational bandwidth, defined as the frequency range over which the responsivity remains within 3 dB of its value at zero frequency, is measured to be 66 kHz. 

The accelerometer's resolution is quantified by its noise equivalent acceleration (NEA), which is influenced by three fundamental noise sources in the optical detection scheme: laser shot noise, detector noise, and thermomechanical noise. The power spectral density (PSD) is a measure of the photodiode output under zero acceleration. The NEA is calculated by normalizing the PSD with the acceleration sensitivity. The NEA is plotted in Figure~\ref{fig:fig3-Setup-Sensitivity-AllanDev}d along with theoretically predicted NEAs for each of the fundamental noise sources. From 20 to 66 kHz, the resolution is 4.2 $\mu g/\sqrt{\mathrm{Hz}}$. Due to a large insertion loss at the waveguide fiber interface (produced by fabrication nonidealities), it was necessary to use an EDFA to reach the shot noise limit for the system (see details in Appendix B.3). For frequencies below 20 kHz, the EDFA introduces additional frequency-dependent noise in the PSD measurements.

\begin{figure*}[htbp]
\centering
\includegraphics[width=5 in]{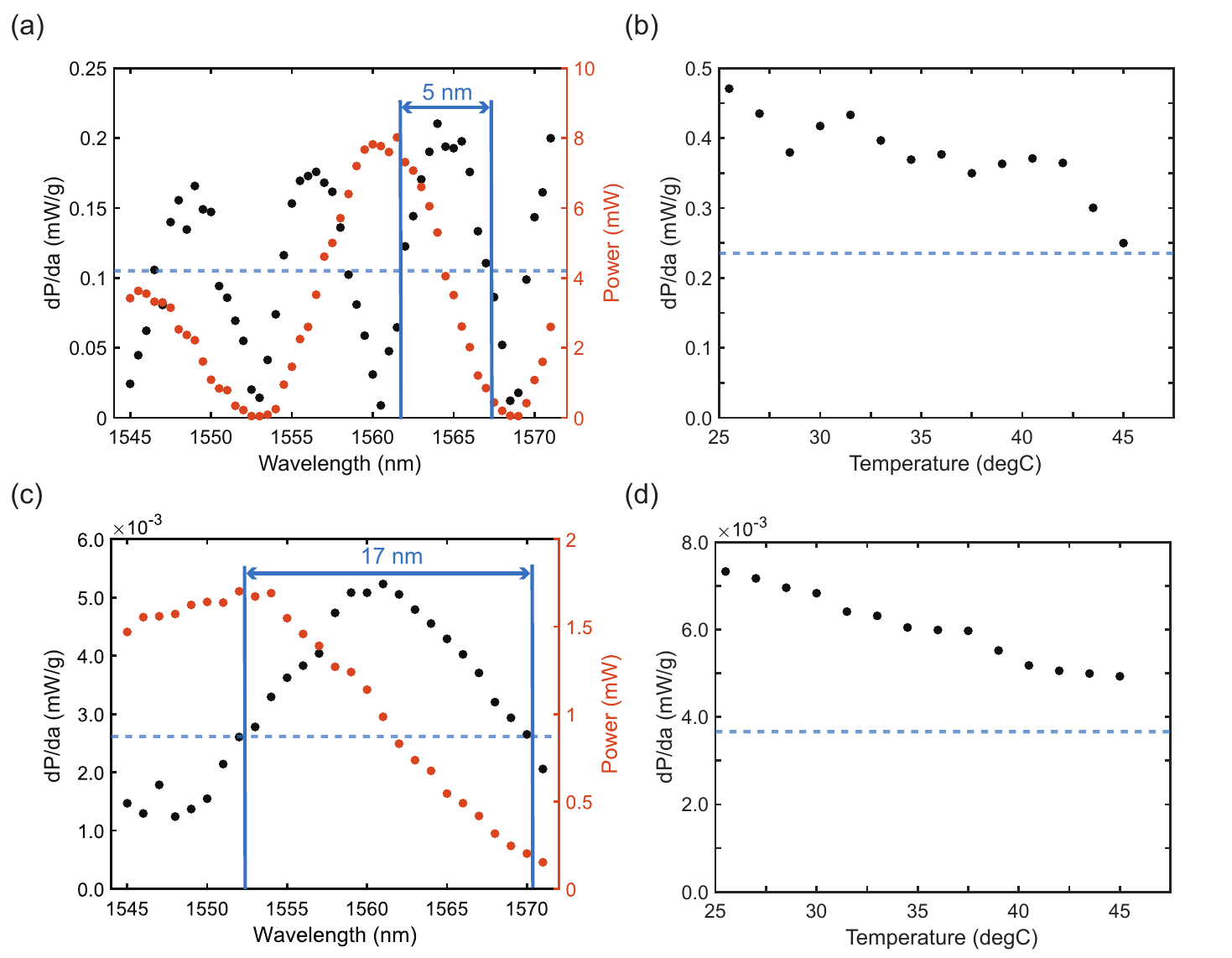}
\caption{\label{fig:fig5} 
Wavelength and temperature sensitivity plots, with the dashed blue line indicating a 3 dB drop in scale factor. (a, b) 4 pass device; (c, d) 2 pass device. }
\end{figure*}

To test the wavelength sensitivity, the temperature of the accelerometer chip was stabilized at 32 $^\circ\mathrm{C}$, while the input wavelength was scanned from 1545 to 1575 nm. Data were collected at 0.5 nm intervals while driving the device at a frequency of 40 kHz, chosen to be well within the off-resonant regime of the device. As shown in Figure~\ref{fig:fig5}a, our device achieves a peak scale factor of 0.21 mW/g at a wavelength of 1564 nm. The sensitivity drops by 3 dB to 0.105 mW/g at +/- 2.5 nm. We also packaged and tested a second device with only two waveguide passes per sensing tether instead of four as the device described in section 2. The 2 pass device achieves a maximum scale factor of 5.2 $\times10^{-3}$ mW/g at a wavelength of 1561 nm, dropping by 3 dB to 2.6 $\times10^{-3}$ mW/g at +/- 8.5 nm (Figure~\ref{fig:fig5}c). 

Separately, we characterized the power transmission of the DSMZIs, revealing wavelength-dependent fringes indicative of an effective path-length mismatch between the arms. The unbalanced DSMZI length is calculated using the free spectral range (FSR), central wavelength $\lambda$, and the effective refractive index ($n_{eff}$) derived from the analytical model. $\delta L = \lambda^2/(FSR \cdot n_{eff})$. Analyzing the period of the wavelength-dependent fringes, we estimate that the device has an effective path length mismatch of 121.4 (27.9) $\mu m$ for the 4 (2) pass devices. The mismatch likely arises from minor variations in the width, thickness, or refractive index of the waveguides, caused by fabrication imperfections. This path length mismatch results in some undesired wavelength sensitivity of the device and an ideal operating wavelength that differs from the design wavelength of 1550 nm. The wavelength dependence of our devices shows a direct correspondence with the phase-dependent output power curves as a function of wavelength, as can be seen in Figure\ref{fig:fig5} a \& c. The responsivity of the device is proportional to the derivative of the transfer function and the bandwidth of the responsivity around the peaks is determined by the FSR of the MZI. In other words, the sensitivity drops off as the effective path length mismatch results in a phase shift away from the most sensitive part of the phase curve. This implies that improvements in the fabrication process would further improve the wavelength insensitivity of our device.

To test the temperature sensitivity of the device, the accelerometer temperature was stabilized at 1.5 $^\circ\mathrm{C}$ intervals over a temperature range of 25.5 $^\circ\mathrm{C}$ to 45 $^\circ\mathrm{C}$. Data were collected while shaking the device at an off-resonance frequency of 40 kHz. As shown in Figure~\ref{fig:fig5}b, the sensitivity of the device drops from  0.47 mW/g at 25.5 $^\circ\mathrm{C}$ to 0.25 mW/g at 45 $^\circ\mathrm{C}$, a factor of 1.88 change. The 2 pass device was also tested and shown to drop from a sensitivity of 7.3 $\times10^{-3}$ mW/g at 25 $^\circ\mathrm{C}$ to 4.9 $\times10^{-3}$ mW/g at 45 $^\circ\mathrm{C}$, a factor of 1.49 change (Fig.~\ref{fig:fig5}d). The temperature sensitivities are likely to be caused by the same effective path length mismatch that causes the wavelength sensitivities, however due to the already small scale of the devices the difference in the path length mismatch that can be induced by temperature is relatively small, and we are therefore unable to see any full fringe patterns over the 20 $^\circ\mathrm{C}$ temperature range that we are able to scan.

The bias instability was found by measuring the Allan deviation with no acceleration applied to the chips. A fixed wavelength high-power laser was used. We measured the output power of the device on a photodiode over the span of 2.5 hours at a sampling rate of 10 kHz. The measurements were conducted in air at ambient temperature, revealing a bias instability of 20.3 (6.3) $\mu g$ at an integration time of 579.3 (234.4) seconds for the 4 (2) pass device. In the Allan deviation plot shown on a log-log scale, Figure~\ref{fig:fig3-Setup-Sensitivity-AllanDev}e, the dotted green line shows the velocity random walk corresponding to the white noise region with a slope of -0.5 (or $\tau^{-0.5}$) (see other noise parameters in Appendix F.3). Our device achieves the lowest bias instability among all optomechanical accelerometers reported in the literature. A comparison with other optomechanical accelerometers that reported bias instability values is summarized in table \ref{Table1}. 
\section{Discussion} \label{sec:DiscussionOutlook}
This work provides a significant step forward for integrated optomechanical accelerometry by moving beyond the traditional regime of small proof masses and high-Q cavities. Using a differential strain-sensing Mach Zehnder interferometer (DSMZI) on the surface of a bulk micromachined proof mass (BMPM), our inertial sensing architecture achieves an acceleration resolution as low as 4.2 $\mu g/\sqrt{\mathrm{Hz}}$ over a 66 kHz bandwidth, coming within a factor of 45 of the best resolution reported for an optomechanical accelerometer, while offering an operational bandwidth that is more than ten times wider. Additionally, we demonstrate that the DSMZI is capable of operating in a substantially larger temperature range ($>$ 20 $^\circ\mathrm{C}$), than previously reported devices, and an optical bandwidth of 17 nm, over 1000 times the optical bandwidth seen in resonator-based devices. We expect even greater temperature and optical bandwidth ranges with improved fabrication to decrease the effective path length difference in the MZI. Our architecture also has, to the best of our knowledge, the lowest bias instability reported for an optomechanical accelerometer, with 6.3 $\mu g$ at an integration time of 243.4 seconds. Due to the large optical bandwidth, the device can be deployed as a system without the need for frequency stabilized or wavelength tunable lasers, making it much more cost-effective and suitable for deployment. The high bandwidth of our design, with the potential to push the same design towards MHz bandwidths, makes it a suitable candidate for applications with large vibrations and impacts, including shock testing, sports science, and condition-based monitoring~\cite{murphy2020choosing,tandon2003detection,espinosa2015inertial}. 
Our current off-resonance NEA is 4.2 $\mu g/\sqrt{\mathrm{Hz}}$. With an optimized coupling scheme, increased waveguide passes, and the implementation of balanced detection using both output ports, the noise floor could potentially be reduced to the thermal noise limit of 130~$\text{ng}/\sqrt{\mathrm{Hz}}$, 330~$\text{ng}/\sqrt{\mathrm{Hz}}$, and 31~$\mu\text{g}/\sqrt{\mathrm{Hz}}$ for the 47.5~kHz, 105~kHz, and 1~MHz designs, respectively. 
These devices have bandwidths that can be made in the 100's of kHz and MHz regimes, which is commensurate with the bandwidths in which substantial vacuum squeezing can generally be produced~\cite{mehmet2011squeezed}. Having low-loss silicon nitride photonics on the same chip, they could thus eventually be integrated with four-wave-mixing-based squeezed light sources to achieve acceleration resolution below the standard quantum limit~\cite{xia2023entanglement}.

Beyond accelerometry, this sensing architecture can potentially be extended to the development of GPS-free, navigation-grade optomechanical vibratory Coriolis gyroscopes by integrating piezoelectric force actuators to drive motion orthogonal to the sense mode. Such a system would enable deployable rotation rate sensing with state-of-the-art bias instability. The design principles developed here achieve ultra-low bias instability, a wide operating temperature range, and a broad optical bandwidth. They can be applied to enhance the performance and deployability of other optomechanical sensors, such as magnetometers and force sensors. This paves the way for optomechanical sensing technologies to have a wider impact in both research and industry. 
\section*{Acknowledgements} \label{sec:acknowledgements}
This work was performed, in part, at the Center for Integrated Nanotechnologies, an Office of Science User Facility operated for the U.S. Department of Energy (DOE) Office of Science. Sandia National Laboratories is a multimission laboratory managed and operated by National Technology \& Engineering Solutions of Sandia, LLC, a wholly owned subsidiary of Honeywell International Inc., for the U.S. Department of Energy’s National Nuclear Security Administration under contract DE-NA0003525. This paper describes objective technical results and analysis. Any subjective views or opinions that might be expressed in the paper do not necessarily represent the views of the U.S. Department of Energy or the United States Government.

\bibliography{references} 
\clearpage
\onecolumngrid

\definecolor{groupA}{RGB}{255, 230, 230}  
\definecolor{groupB}{RGB}{230, 255, 230}  
\definecolor{groupC}{RGB}{230, 230, 255}  
\definecolor{groupD}{RGB}{255, 255, 200}  
\begin{table*}[t]
\centering
\caption{Summary of the specifications of optomechanical accelerometers reported in the literature}
\label{Table1}
\renewcommand{\arraystretch}{1.5}
\begin{tabular}{|p{2.2cm}|p{1.5cm}|p{0.9cm}|p{1.4cm}|p{1.8cm}|p{2cm}|p{0.9cm}|p{1.6cm}|}
    \hline
    Paper& Mass (mg) &$f_s$ (kHz) & Q-factor& Operational Bandwidth (kHz)& Off-resonance Resolution ($\mu g/\sqrt{\mathrm{Hz}}$)& BI ($\mu g$)& Integration time (sec) \\ \hline

    \rowcolor{groupD}
    Li~\cite{li2018characterization}&5.7$\times 10^{-2}$& 0.13& 44.3 & 0.1 & 4.5 & 31.8&1 \\ \hline


    \rowcolor{groupB}
    Zhou~\cite{zhou2021broadband}&20&8.74& 66, 565* & 6.2 & 0.093, 0.032* & N/A&N/A \\ \hline

    \rowcolor{groupB}
    Gerberding~\cite{gerberding2015optomechanical}&25& 10.6& 1.2$\times 10^4$* & 7.5 & 8.0* & 295.7*& 2$\times 10^{-3}$ \\ \hline

    \rowcolor{groupB}
    Guzman ~\cite{guzman2014high}&25& 10.7 & 4.2$\times 10^4$* & 7.6 & 0.1* & N/A&N/A \\ \hline

    \rowcolor{groupB}
    Long~\cite{long2023high}&11.1&23.7& 115 & 16.7 & 8 & 305.9& 1.9$\times 10^{-3}$ \\ \hline

    \rowcolor{groupC}
    Krause~\cite{krause2012high}&1$\times 10^{-5}$& 27.5& 1.4$\times 10^6$* & 19.4 & 10* & N/A&N/A \\ \hline

    \rowcolor{groupC}
    Huang~\cite{huang2020chip}&5.6$\times 10^{-3}$& 71.3& 1383* & 50.4 & 8.2* &50.9*& 6$\times 10^{-2}$ \\ \hline

    \rowcolor{groupA}
    Our devices&16.7& 93.4 (93.3)& 54.3 & 66 & 4.2 \newline (29) & 20.3 (6.3)& 579.3 (243.4) \\ \hline
\end{tabular}

\vspace{1mm}
The state-of-the-art optomechanical accelerometers summarized in the table are categorized into the following groups based on their optical transduction mechanisms: (1) Microsphere-based resonant accelerometer (highlighted in yellow), (2) Fabry–Pérot microcavity-based resonant accelerometers (in green), (3) photonic crystal cavity-based resonant accelerometers (in blue), and (4) integrated Mach–Zehnder interferometer (MZI)-enabled strain-sensing accelerometers (in red). Devices are listed in order of increasing mechanical sensing bandwidth.\\
Characterization is performed in vacuum*
\end{table*}
\clearpage
\appendix 
\section{Harmonic oscillator susceptibility} \label{sec:HarmonicOscillatorSusceptibility}

Our optomechanical accelerometer employs a single bulk micromachined proof mass (BMPM) suspended by tethers to form a 1 Degree-of-Freedom (DoF) harmonic oscillator. The motion of the oscillator can be described by the Langevin equation.

\begin{equation} \label{eq:S1}
m \frac{d^2 x(t)}{dt^2} +c \frac{dx(t)}{dt} + k x(t) = F_{\text{ext}}(t).
\end{equation}
Where m is the effective mass, x is the displacement of the mass, c is the damping coefficient, k is the spring constant, and $F_{ext}$ is the external force applied to the oscillator. 
In the frequency domain, this equation becomes 
\begin{equation} \label{eq:S2}
-\Omega^2 x +i c \Omega x +{\Omega_s}^2 = \frac{F_{\text{ext}}(\Omega)}{m}.
\end{equation}

The mechanical resonance frequency of the sense mode is $\Omega_s = \sqrt{k/m} $.
The displacement of the proof mass x($\Omega$) is proportional to the mechanical susceptibility \begin{equation} \label{eq:S3}
\chi(\Omega) = \frac{x(\Omega)}{a(\Omega)} = \frac{1}{\Omega_s^2 - \Omega^2 + \frac{i \Omega_s \Omega}{Q}}.
\end{equation}
Where $Q$ is its mechanical quality factor. 
\section{Noise analysis} \label{sec:Noise Analysis}
\subsection{Noise from thermal Brownian motion}
Thermal noise is one of the fundamental limits to the
precision of mechanical measurements. The $F_{th}$ is a random force with a white spectrum density added to $F_{ext}$ shown in (\ref{eq:S1}), it can be expressed as
\begin{equation} \label{eq:S4}
{F_{th}}^2 = 4k_BTf.
\end{equation}
Where $k_B$ is Boltzmann's constant, T is temperature, and f is a Gaussian white noise. The noise from thermal Brownian motion derived from fluctuation-dissipation theorem \cite{saulson1990thermal} is
\begin{equation} \label{eq:S5}
a_{th} = \sqrt{S^{th}_{aa}} = \sqrt{\frac{4k_BT\Omega_s}{mQ}}.
\end{equation}
\subsection{Shot and detector noises}
Shot noise is generated by the dc power hitting the photodiode, it can be expressed by
\begin{equation} \label{eq:S6}
a_{shot} = ({\frac{dP}{da}})^{-1} \sqrt{\frac{2\hbar\omega_0 P_{tot}}{\eta_{qe}}}.
\end{equation}
Where $\frac{dP}{da}$ is the scale factor of the device, $\omega_0$ is the optical frequency of light, and $P_{tot}$ is the total power on the detector. $\eta_{qe}$ is the quantum efficiency and it can be described as $\frac{R\hbar\omega_0}{e}$.
 Hence, shot noise can be written by 

\begin{equation} \label{eq:S7}
a_{shot} = ({\frac{dP}{da}})^{-1} \sqrt{\frac{2 e P_{tot}}{R}}.
\end{equation}
Where $e$ is the electrical charge, $R$ is the responsivity of the photodiode.

Detector noise is quantified by its noise-equivalent-power (NEP) and the optical noise power-spectral-density is 

\begin{equation} \label{eq:S8}
S^{\text{NEP}}_{PP} = {\text{NEP}}^2.
\end{equation}
We use the Newport 2053 detector that has a NEP of 0.34 $pW/\sqrt{\mathrm {Hz}}$. The noise induced by the detector's NEP is
\begin{equation} \label{eq:S9}
a_{det} = ({\frac{dP}{da}})^{-1} \text{NEP}.
\end{equation}

\subsection{Optical power spectral density measurements} \label{SI:2.3}
We use a lock-in amplifier to measure the photodiode output voltage in $V_{rms}$ with a time constant of 100 ms and an impedance $Z$ of 50 $\Omega$.
The Power Spectral Density (PSD) in dBm/Hz can be expressed as
\begin{equation} \label{eq:S10}
\text{PSD} = 10 \log_{10}\left( \frac{V_{\text{rms}}^2}{Z} \times 1000 \right) - 10 \log_{10}(\text{BW}).
\end{equation}

\begin{figure}[h!]
\centering
\includegraphics[width=0.6\linewidth]{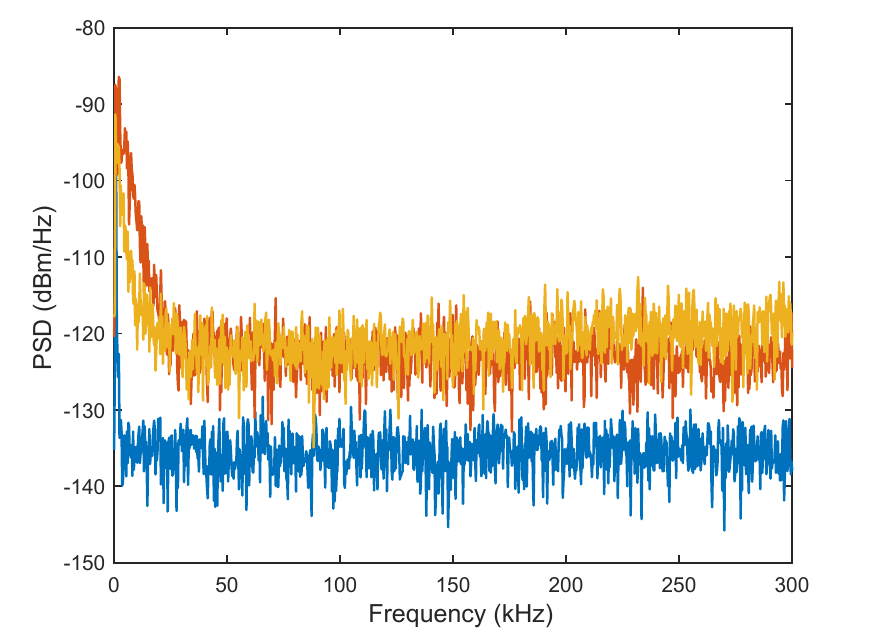}
\caption{\label{fig: PSD_comp} The comparison of Power Spectral Density (PSD) measurements with various input sources. The red curve represents the 1550 nm laser light amplified to 750 mW using an erbium-doped fiber amplifier (EDFA). The blue curve corresponds to the 1550 nm laser light without amplification, producing an output power of 10 mW. The yellow curve displays the PSD measurement taken with a fixed-wavelength, low-noise laser at 1550 nm with an output power of 750 mW.}
\end{figure}

The noise-equivalent-acceleration (NEA) measurement in Fig. \ref{fig:fig3-Setup-Sensitivity-AllanDev}d from the main text reveals that the low-frequency performance is dominated by 1/f noise. Fig.\ref{fig: PSD_comp} presents the PSD of the optically transduced signal from our device for three different input light sources: an erbium-doped fiber amplifier (EDFA)-boosted light at 750 mW (red), a 1550 nm laser at 10 mW without an EDFA (blue), and a fixed-wavelength, low-noise 1550 nm laser at 750 mW (yellow).

\section{Modeling strain-induced phase variation in the waveguide} \label{sec:Modeling strain-induced}
The DSMZI measures acceleration by detecting the displacement of the proof mass, which induces strain and deformation in the waveguides embedded in the tethers. To sense these changes, we analyze the transfer function of the optical power at the two output ports of the interferometer. The optomechanically induced change in phase due to the strain and deformation induced by the acceleration can be expressed as: 
\begin{equation} \label{eq:3}
\phi_{\pm} = \Delta(kL)_{\pm}(a) = k\Delta L(a)\pm \Delta (kL)_{PE}\pm \Delta (kL)_{MB}.
\end{equation}
The first term $k\Delta L(a)$ accounts for the change in the total optical path length within the strained tether. It is expressed as: 
\begin{equation} \label{eq:4}
k\Delta L(a) = k(L \pm \chi a).
\end{equation} 
Where $k$ is the propagation constant, $L$ is the physical length of the waveguides, $\chi$ is the mechanical susceptibility ($dx/da$), and $\chi a$ is the displacement of the proof mass. Under zero acceleration ($a=0$), the optical phase shift $\phi$ simplifies to $kL$, resulting in $\phi_\Delta = 0$. The second term, $\Delta (kL)_{PE}$, describes the change in the accumulated phase due to the photoelastic change in the propagation constant, which arises from strain-induced modifications in the refractive index. Since the waveguide experiences a non-uniform strain along the tether, $\Delta k$ varies as a function of $x$. The accumulated phase shift due to the photoelastic effect is obtained by integrating $\Delta k_{PE}$ along $x$ from 0 to L: 
\begin{equation} \label{eq:SI kL_PE}
 \Delta (kL)_{PE} = \int_{0}^{L}{dx} \Delta k_{PE}(x;a) = \frac{\omega}{2}(\frac{d\omega}{d\beta})^{-1}\int_{0}^{L}{dx}\frac{ \iint{dA}\ \mathbf{E}^* \cdot \Delta \epsilon_{pe}(x;a) \cdot \mathbf{E}}{max|\mathbf{u}(x)|\iint{dA\ \mathbf{E}^* \cdot \epsilon \cdot \mathbf{E}}}.
\end{equation}
Where $\frac{d\omega}{d\beta}$ is the group velocity, $\Delta \epsilon_{pe}(x;a)$ is the photo-elastic tensor perturbation in the permittivity, and $max|\mathbf{u}(x)|$ is the maximum of the mechanical displacement field at any point along the tether \cite{tian2024piezoelectric,chen2006appendix}.

Lastly, $\Delta (kL)_{MB}$ is the phase shift due to the opto-mechanical moving boundary effect, which induces a change in propagation constant due to waveguide dispersion with an altered cross-section by mechanical displacement \cite{johnson2002perturbation}:

\begin{equation} \label{eq:SI kL_MB}
 \Delta (kL)_{MB} = \int_{0}^{L}{dx} \Delta k_{MB}(x;a) = \frac{\omega}{2}(\frac{d\omega}{d\beta})^{-1}\int_{0}^{L}{dx}\frac{ \oint\,dl (\mathbf{u}(x) \cdot \hat{\mathbf{n}})[\Delta \epsilon |\mathbf{E}_{\parallel}|^2-\Delta \epsilon^{-1}|\mathbf{D}_{\perp}|^2]}{max|\mathbf{u}(x)|\iint{dA\ \mathbf{E}^* \cdot \epsilon \cdot \mathbf{E}}}.
\end{equation}
$\Delta \epsilon$ is the difference in permittivity and $\Delta \epsilon^{-1}$ is the difference in the reciprocal of the permittivity between the waveguide core and cladding.

The final differential phase, $\phi_\Delta$, obtained by subtracting one arm’s accumulated phase by the other’s can be expressed as:
\begin{equation} \label{eq:7}
\phi_\Delta = \phi_{+}-\phi_{-}=2[k\cdot (\chi a)+\Delta (kL)_{PE}+ \Delta (kL)_{MB}].
\end{equation}
A 3D finite element method (FEM) model is used to evaluate the integrals in Eq.(\ref{eq:SI kL_PE},\ref{eq:SI kL_MB}) numerically using the exact strain profiles induced by acceleration of the proof mass. When the accelerometer is driven with a 1 g acceleration at a frequency substantially less than the resonance frequency, the proof mass is expected to displace by 22.5 pm along the sense mode direction. Fig. \ref{fig:fig2}a from the main text shows its displacement field, $\mathbf{u}(x,y,z)$, when applying a prescribed displacement to the proof mass with a magnitude of 22.5 pm. This causes equal and opposite strain and deformation in the sensing tethers. In the model, boundary conditions are used to launch light from one end of the waveguide and collect it in the other, calculating the scattering matrix between the two boundaries, which in turn allows extracting the accumulated phase. The two inset figures in Fig. 2a depict the mode profiles in the oppositely deformed and strained waveguide cross sections, causing equal and opposite phase accumulation for the TE waveguide modes in each arm. To account for the photoelastically induced phase shift, $\phi_\Delta$, described by Eq.(\ref{eq:SI kL_PE}), the refractive indices of the waveguide core (\(\text{SiN}_x\)) and cladding (\(\text{SiO}_2\)) are modified by the photoelastic effect, given by: 
{$\Delta\frac{1}{ n_{i}^2} = \sum_{j=1}^{6} p_{ij}S_{j}$\cite{chen2006appendix,tian2024piezoelectric}.} Here, \(p_{ij}\) is the photoelastic tensor, and \(S_{j}\) is the strain tensor. 

The finite element model incorporates a deformable mesh that responds to the displacement of the proof mass, which allows the model to directly calculate changes in physical path length and moving boundary optomechanical coupling. A scattering matrix is used to calculate the total differential phase accumulation between the two interferometer arms, resulting in a phase change of $\phi_\Delta = 5.76 \times 10^{-4}~\text{rad}/g$. This analytical model enables us to predict the sensor’s acceleration-to-power responsivity and noise-equivalent acceleration (NEA), while accounting for thermal, shot, and detector noise. Under the assumption that 10 mW of optical power is hitting each detector with a balanced detection scheme, the sensor is expected to achieve a displacement resolution up to 0.03 $fm/\sqrt{\mathrm{Hz}}$ without the use of an optical resonator, corresponding to an off-resonance shot-noise limited acceleration sensitivity of 1.15 $\mu g/\sqrt{\mathrm{Hz}}$ (See Appendix D).
\section{Theoretical noise-equivalent-acceleration analysis} \label{sec:NEA}
The optical power at one of the output ports of the DSMZI can be expressed as
\begin{equation}
P = P_0 \sin^2\left(\frac{\phi_\Delta+\pi /2}{2}\right),
\end{equation}
where $P$ is the output power, $P_0$ is the input power, and $\phi_\Delta$ is the differential phase shift between the two interferometer arms. 

The scale factor of the accelerometer, defined as the change in output power per unit acceleration, is given by
\begin{equation} \label{eq:S17}
\frac{dP}{da} = \frac{dP}{d\phi} \cdot \frac{d\phi}{da} = P_0 \sin(\frac{2\phi_\Delta+\pi}{4}) \cos(\frac{2\phi_\Delta+\pi}{4})\cdot \frac{d\phi}{da},
\end{equation}
where $\frac{d\phi}{da}$ quantifies the phase shift induced by strain and deformation in response to applied acceleration. 

In a balanced detection scheme, the scale factor is effectively doubled, and common-mode noise such as laser intensity fluctuations can be significantly suppressed. Assuming that the DSMZI is perfectly balanced, the transfer function exhibits its maximum slope, thus maximizing the interferometric sensitivity. Under this condition, the scale factor simplifies to
\begin{equation}
\frac{dP}{da} = P_0 \cdot\frac{d\phi}{da}.
\end{equation}
Using the analytical model described in the accelerometer design section, the phase accumulation difference between the two DSMZI arms is $\frac{d\phi}{da} = 5.76 \times 10^{-4}~\text{rad}/g$ per waveguide pass. For a high-sensitivity device with four waveguide passes per sensing tether and an incident optical power of 10~mW on each photodetector, the resulting scale factor $\frac{dP}{da}$ is $5.76\times 10^{-5}~\text{W}/g$ in the flat frequency response region.
The inverse of the scale factor can then be used to predict the shot and detector noise, see Equations~(\ref{eq:S7}) and (\ref{eq:S9}). By modifying the proof mass size and tethers dimensions, we explore designs with a wide range of mechanical frequencies, from $<$ 50 kHz to 1 MHz. In Table \ref{Table S1}, we list three designs with their resonance frequency, effective mass, theoretical off-resonance thermal, shot, detector noise, and noise-equivalent-acceleration (NEA). The effective mass of the proof mass is derived using this expression
\begin{equation} \label{eq:S15}
m_{\text{eff}} = \frac{1}{x^2} \int dV \, \rho |Q(x)|^2,
\end{equation} 
 where $\rho$ is the mass density and $Q(x)$ is the displacement field.
\begin{table}[h]
\centering
\renewcommand{\arraystretch}{1.5}
\begin{tabular}{p{1.6cm}p{1.7cm}p{2.3cm}p{2.5cm}p{2.4cm}p{2.6cm}}

      $f_s$ & $m_{eff}$ (mg)  & $a_{th}$ ($\mu g/\sqrt{\mathrm{Hz}}$) & $a_{shot}$ ($\mu g/\sqrt{\mathrm{Hz}}$)& $a_{det}$ ($\mu g/\sqrt{\mathrm{Hz}}$) & NEA ($\mu g/\sqrt{\mathrm{Hz}}$)\\ \hline
    47.5 kHz & 56.0 & 0.130 & 0.224  & 0.00121 & 0.259 \\ \hline
    105 kHz & 16.7 & 0.354 & 1.10 & 0.00590 & 1.15\\ \hline
    1 MHz& 1.24 & 4.03 & 31.0 & 0.167 & 31.3\\ \hline

\end{tabular}
\caption{Simulated properties of three designs with resonance frequencies ranging from below 50~kHz to 1~MHz, achieved by varying the proof mass size and tether dimensions. The estimated contributions of different noise sources to the noise-equivalent acceleration (NEA) in the off-resonance, flat frequency response region are also shown. The frequency-dependent behavior of these noise sources is plotted in Fig.~2 of the main text.}

\label{Table S1}
\end{table}

\section{Full experimental characterization of our devices} \label{sec:Experimental Characterization}
The full characterization of both devices mentioned in the main text is summarized in Table \ref{Table S1}. Since the mechanical structures are identical, both devices theoretically share the same fundamental mechanical resonance frequency. The key intrinsic difference lies in the number of waveguide passes on the sensing tethers: the 4 pass device is designed to have twice the scale factor of the 2 pass device, resulting in an off-resonance shot noise-limited NEA that is half as large.

Several approaches can be employed to further reduce the bias instability. The devices can be characterized using a measurement setup that offers enhanced thermal and vibration isolation. Laser intensity can be stabilized using an electro-optic modulator, while polarization drift can be mitigated by replacing single-mode (SM) fibers with polarization-maintaining (PM) fibers. Improved balancing of the MZI arms is also expected to lead to an improvement in bias instability. 

Imperfections in the fabrication process led to insertion losses on the order of 12 dB per facet in our accelerometer chips. Due to these high coupling losses, we had to use an erbium-doped fiber amplifier (EDFA) to amplify the input light to 1 W at the input of our chip in order to get around 4 mW of power at the detectors. We expect to be able to reduce the losses in future designs to around 1.5 dB per facet, reducing the required input power to less than 10 mW for results comparable to those reported here. Furthermore, NEA measurement in Figure~\ref{fig:fig3-Setup-Sensitivity-AllanDev}d shows that the off-resonance performance is limited by \(1/f\) noise from the EDFA, therefore we also expect to see an improvement in NEA with reduced facet losses. 

\begin{table}[h]
\centering
\renewcommand{\arraystretch}{1.5}
\begin{tabular}{p{2.4cm}p{1.2cm}p{1.8cm}p{1.9cm}p{1.8cm}p{1.9cm}p{1.5cm}}

      & $f_s$ \newline (kHz)  & $NEA_{off}$ \newline ($\mu g/\sqrt{\mathrm{Hz}}$) & Unbalanced length ($\mu m$) & Wavelength range ($nm$)& Temperature range ($^\circ\mathrm{C}$) & Bias instability \\ \hline
    4 pass device & 93.4 & 4.2 & 121.4  & 5 & $>$ 20  & 20.3 $\mu g$ at 579.3 s \\ \hline
    2 pass device & 93.3 & 29 & 27.9 & 17 & $>$ 20 & 6.3 $\mu g$ at 243.4 s \\ \hline

\end{tabular}
\caption{Measured performance comparison of two devices}
\label{Table S2}
\end{table}

\section{Comparisons with state-of-the-art optomechanical accelerometers} \label{sec:Comparisons}
\subsection{Optical bandwidth analysis}
For cavity-based optomechanical accelerometers, the 3 dB optical bandwidth is equivalent to the optical full width at half maximum (FWHM). In the case of Fabry–Pérot microcavity accelerometers, the FWHM can be directly calculated as the free spectral range (FSR) divided by the cavity finesse. In Table IV, we expand the comparison presented in the main text and summarize the corresponding optical bandwidths of the devices reported previously.

\definecolor{groupA}{RGB}{255, 230, 230}  
\definecolor{groupB}{RGB}{230, 255, 230}  
\definecolor{groupC}{RGB}{230, 230, 255}  
\definecolor{groupD}{RGB}{255, 255, 200}  

\begin{table}[htbp]
\centering
\label{Table optical BW}
\renewcommand{\arraystretch}{1.5}
\begin{tabular}{|p{2.25cm}|p{0.95cm}|p{1.7cm}|p{2.1cm}|p{1.0cm}|p{1.7cm}|p{1.7cm}|}

    \hline
    Paper& $f_s$ (kHz) & Operational Bandwidth (kHz)& Off-resonance Resolution ($\mu g/\sqrt{\mathrm{Hz}}$)& BI ($\mu g$)& Integration time (sec)&Optical Bandwidth (nm) \\ \hline
 
    \rowcolor{groupD}
    Li~\cite{li2018characterization}& 0.13 & 0.1 & 4.5 & 31.8&1&2.5$\times 10^{-4}$ \\ \hline

    \rowcolor{groupB}
    Zhou~\cite{zhou2021broadband}&8.74 & 6.2 & 0.093, 0.032* & N/A&N/A&5.9$\times 10^{-4}$ \\ \hline

    \rowcolor{groupB}
    Long~\cite{long2023high}&23.7 & 16.7 & 8 & 305.9& 1.9$\times 10^{-3}$&1.2$\times 10^{-3}$ \\ \hline

    \rowcolor{groupB}
    Gerberding~\cite{gerberding2015optomechanical}& 10.6 & 7.5 & 8.0* & 295.7*& 2$\times 10^{-3}$&4.4$\times 10^{-3}$ \\ \hline

    \rowcolor{groupB}
    Guzman~\cite{guzman2014high}& 10.7 & 7.6 & 0.1* & N/A&N/A&4.4$\times 10^{-3}$ \\ \hline

    \rowcolor{groupC}
    Krause~\cite{krause2012high}& 27.5 & 19.4 & 10* & N/A&N/A&1.6$\times 10^{-1}$ \\ \hline

    \rowcolor{groupC}
    Huang~\cite{huang2020chip}& 71.3 & 50.4 & 8.2* &50.9*& 6$\times 10^{-2}$&3.6$\times 10^{-1}$ \\ \hline

    \rowcolor{groupA}
    Our\newline devices& 93.4 (93.3) & 66 & 4.2\newline(29) & 20.3 (6.3)& 579.3 (243.4)&5\newline(17) \\ \hline
\end{tabular}
\caption{The optomechanical accelerometers are listed in order of increasing optical bandwidth. (* The characterization is performed in vacuum.)}
\end{table}

\subsection{Temperature bound}

For an example of the temperature range that can be expected from a resonator-based optomechanical accelerometer, we follow the temperature analysis given in \cite{krause2012high}.
The laser cavity detuning induced by a change in temperature is
\begin{align}\label{eq:S20}
\Delta_{th} = g_{th}\Delta T_o,
\end{align}
where $g_{th}$ is the thermo-optical tuning coefficient.
\begin{align}
g_{th} = \frac{dn}{dT} \frac{\omega_c}{n}.
\end{align}
$\frac{dn}{dT} = \alpha_{TO} + \alpha_{TE}$ is the thermo optic coefficient of and thermal expansion coefficient of the material.
\newline

To find the temperature change corresponding to a 50\% drop in transmission, we set the detuning equal to $\lambda_o\pm \frac{\Delta \lambda}{2}$, where $\lambda_o$ is the central wavelength of the laser and $\Delta \lambda$ is FWHM/2. 

This is equivalent to a detuning of:

\begin{align}
\Delta_{th} = \frac{c}{\lambda_o - \frac{\Delta \lambda}{2}} -  \frac{c}{\lambda_o + \frac{\Delta \lambda}{2}}
\end{align}

We can then solve for $\Delta T_o$ using Eq.(\ref{eq:S20}),
\begin{align}
\Delta T_o = \frac{\Delta_{th}}{g_{th}} = (\frac{c}{\lambda_o - \frac{\Delta \lambda}{2}} -  \frac{c}{\lambda_o + \frac{\Delta \lambda}{2}})(\frac{dn}{dT})^{-1}\frac{n }{\omega_c}
\end{align}

Using $\lambda_o = 1537.36$ nm, $Q_0 = 9500$, $\alpha _{TO} = 2.51 \times 10^{-5}~K^{-1}$~\cite{johnson2022determination}, and $\alpha _{TO} = -3.7 \times 10^{-6}~K^{-1}$~\cite{azom_silicon}, we find $\Delta T_o = 0.69~^\circ\mathrm{C}$.

\subsection{Bias instability and Allan deviation}

\begin{figure}[h!]
\centering
\includegraphics[width=0.6\linewidth]{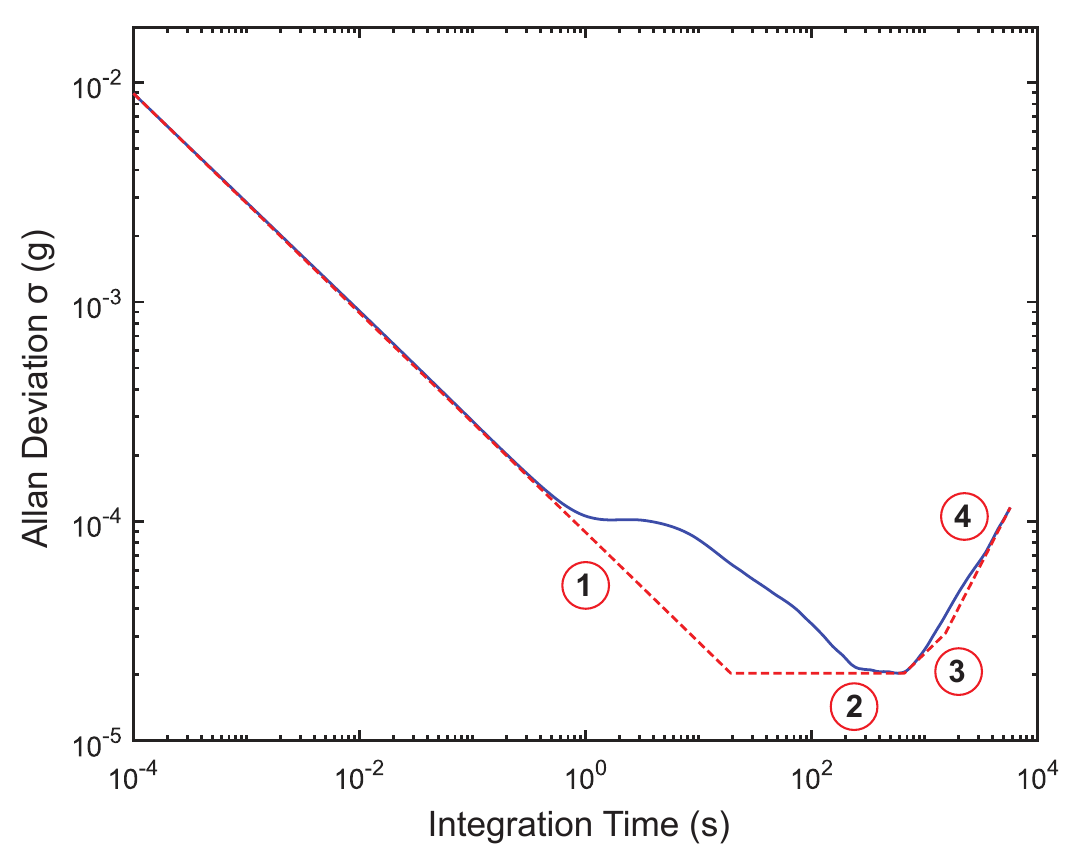}
\caption{\label{fig:AllanFit} Allan deviation $\sigma(\tau)$ as a function of integration time $\tau$, with fitted noise terms: (1) velocity random walk (VRW), (2) bias instability, (3) acceleration random walk (ARW), and (4) rate ramp (RR).}
\end{figure}

To measure bias instability, the accelerometer is placed in an acoustically shielded enclosure under ambient conditions (in air at room temperature) without external acceleration applied. The overlapped Allan deviation is computed using the MATLAB AVAR GUI \cite{ogier2025avar}, which is also used to fit and extract various noise coefficients. In the log–log scale Allan deviation plot shown in Fig.\ref{fig:AllanFit}, the dotted red line in region (1) represents the velocity random walk (VRW) component, corresponding to the white noise region. This region is fitted with a slope of $-0.5$ (i.e., $\tau^{-0.5}$) and has a magnitude of 89.2$\mu g/\sqrt{\mathrm{Hz}}$. The bias instability appears at the minimum of the Allan deviation curve at 579.3 s, where the slope is zero. The acceleration random walk (ARW) and the rate ramp (RR) exhibit slopes of 0.5 and 1, respectively. From this analysis, we extract a bias instability of 20.3~$\mu g$, an ARW of 0.79~$\mu g/\sqrt{\mathrm{s}}$, and an RR of 0.02~$\mu g/\mathrm{s}$. 
\section{Dynamic range analysis} \label{sec:Dynamic range}
From the derivations in section D, the scale factor of a perfectly balanced DSMZI at zero acceleration is 
\begin{equation}
\frac{dP}{da} = P_0 \cdot\frac{d\phi}{da}.
\end{equation}
We want to calculate the upper limit of acceleration at 1\% scale factor deviation that corresponds to the upper limit of the linear dynamic range. Therefore, we need to first solve for $\phi_\Delta$ in the expression below
\begin{equation} \label{eq:S27}
P_0  \cos(\phi_\Delta)\cdot \frac{d\phi}{da}  =  P_0 \cdot\frac{d\phi}{da}\cdot(1-1\%).
\end{equation}
$\phi_\Delta$ is 0.14 in radians. From the analytical model, the phase accumulation difference between the two DSMZI arms is $\frac{d\phi}{da} = 5.76 \times 10^{-4}~\text{rad}/g$ per waveguide pass. For our 4 pass device, the resulting upper limit of acceleration is 61.3 g. The dynamic range is defined as $20\cdot log (a_{max}/a_{min})$, where $a_{max}$  stands for the maximum acceleration based on when the response deviates from linearity, and $a_{min}$ is the minimum detectable acceleration. The expected noise floor of this 105 kHz accelerometer design is 330 $ng/\sqrt{\mathrm{Hz}}$. With a measurement bandwidth of 1 Hz, the expected dynamic range is 165.4 dB (1.86 $\times~10^8$ in ratio). In the experiment, we applied various voltages to drive the shear piezo actuator and observed that our device exhibited a linear response over a dynamic range of 81.6 dB. This measurement was limited by the maximum acceleration induced by the actuator. In future work, we plan to perform high-g measurements using a calibrated shaker table and a commercial reference accelerometer to further characterize the dynamic range. Unlike cavity-based optomechanical accelerometers, our device is mechanically robust and capable of withstanding high-g tests, and its linearity is not constrained by the optical cavity linewidth.  

\section{Motion suppression by the shear piezo actuator} \label{sec:Motion suppression}
\begin{figure}[h!]
\centering
\includegraphics[width=0.9\linewidth]{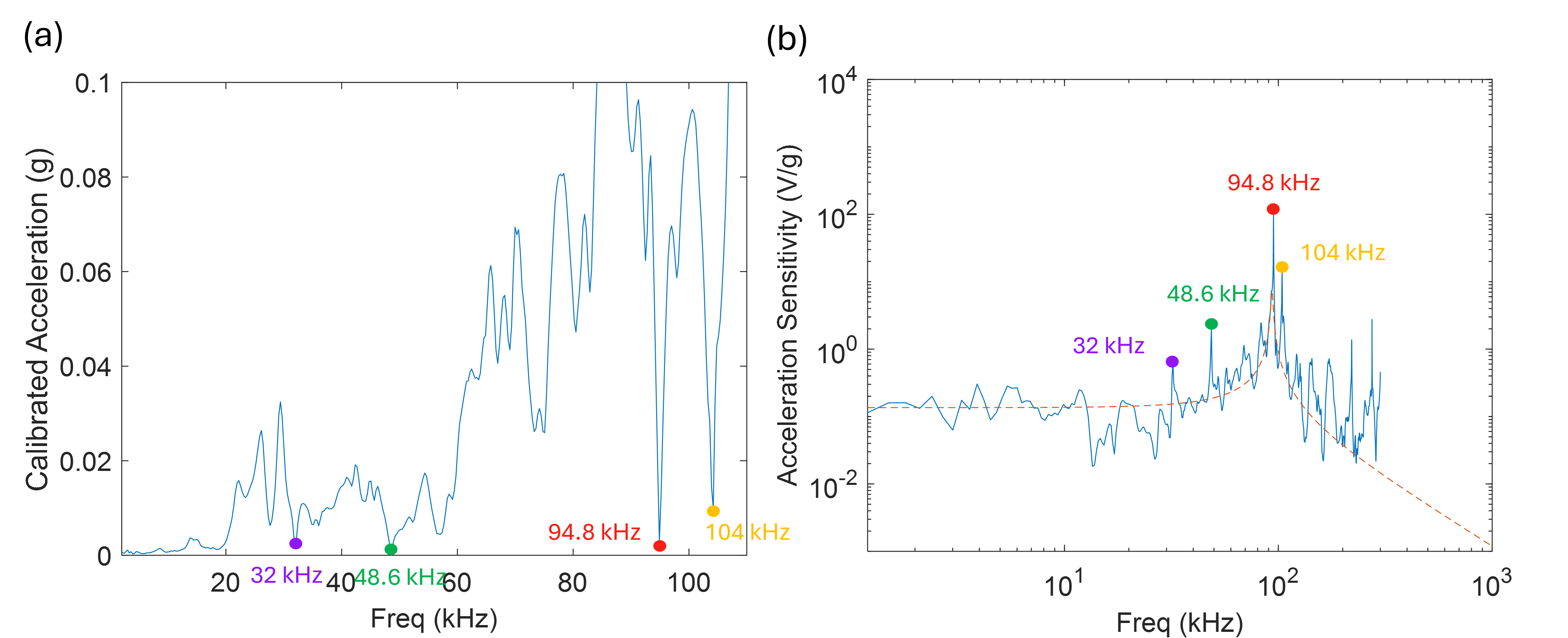}
\caption{\label{fig: Calibrated Accel} Frequency dependence of motion suppression and sensitivity. (a) Applied acceleration as a function of frequency, showing dips where motion is suppressed by the shear piezo actuator. (b) Acceleration sensitivity curve, where the sharp peaks correspond to the suppressed motion frequencies observed in (a).}
\end{figure}

A free-space Michelson interferometer built to calibrate the acceleration applied by the shear piezo actuator is shown in Fig. \ref{fig:fig3-Setup-Sensitivity-AllanDev}a of the main text. The test arm bounces off the edge of the accelerometer chip, and the reference arm passes through an attenuator to match the power reflected from the chip before being reflected off a piezoelectric-actuated mirror. The light recombines on the beam-splitter and is detected by a photodiode.  The piezoelectric-driven mirror is controlled by the PI lock box so that the interferometer’s state is locked at the steepest slope of its transfer function. This filters out low-frequency noise and maximizes the measured response. The shear piezo actuator is driven by a sinusoidal wave generated by the internal source of the lock-in amplifier, The displacement of the shear actuator is described by $x = b\ V_0\ sin(2 \pi ft)$, where b is the piezo voltage displacement constant, $V_0$ is the amplitude of the applied voltage. The acceleration applied to the chip mounted on the shear actuator can be calculated by taking the second derivative of the displacement expression, so the acceleration is $a = -x\ \omega^2$. From the acceleration calibration data, we observed that the displacement of the external actuator varies with frequency, leading to non-quadratic acceleration as a function of frequency, as shown in Fig.\ref{fig: Calibrated Accel}a. The labeled dips indicate motion suppression at specific frequencies. Consequently, this suppression causes the sharp peaks in Fig.\ref{fig: Calibrated Accel}b, which is obtained by normalizing the accelerometer's frequency response to the applied acceleration.

\section{Fabrication} \label{sec:Fabrication}
Fabrication begins with growth of three microns of LPCVD oxide on handle silicon. Immediately following is the growth of $300 nm$ of LPCVD grown high-stress, stoichiometric $SiN_x$. The $SiN_x$ waveguide is patterned via electron beam lithography using an JEOL 100keV system employing a negative tone resist process. Etching of the $SiN_x$ is accomplished via Reactive Ion Etching (RIE) employing both SF\(_6\) and CHF\(_3\) mixed mode chemistry. The wafer is then cleaned using a standard RCA process before three microns of a silane based PECVD oxide cladding is deposited on the surface of the wafer. Optical contact lithography is employed in order to pattern etch windows and fiber trenches on the frontside of the wafer. The windows are etched into the oxide and partially into the bulk $Si$ handle wafer using Inductively Coupled Plasma (ICP-RIE) etching following a standard, high power SF\(_6\) based chemistry. Once the frontside has been patterned and etched, contact lithography is once again employed in order to pattern the proof mass definition on the backside of the wafer. Front-to-back alignment of a 40 $\mu m$ thick resist is accomplished using a KarlSuss MA6. Through wafer etching is accomplished using a Deep Reactive Ion Etcher (DRIE) employing a standard Bosch chemistry. The backside through etching meets the frontside head start etch partially through the wafer, reducing lateral blowout of the features that tends to accompany over etching with typical Bosch processes.
\end{document}